\pdfoutput=1
\documentclass[11pt,a4paper]{article}

\usepackage[utf8]{inputenc}
\usepackage[T1]{fontenc}
\usepackage{lmodern}
\usepackage[margin=1in]{geometry}
\usepackage{amsmath,amssymb,amsthm}
\usepackage{graphicx}
\usepackage{booktabs}
\usepackage{xcolor}
\usepackage{natbib}
\usepackage{hyperref}
\usepackage{cleveref}
\usepackage{enumitem}
\usepackage{tikz}
\usetikzlibrary{shapes.geometric,arrows.meta,positioning,fit,calc}
\usepackage{caption}
\usepackage{multirow}
\usepackage{makecell}
\usepackage{mdframed}

\hypersetup{
  colorlinks=true,
  linkcolor=blue!70!black,
  citecolor=green!50!black,
  urlcolor=blue!60!black,
}

\newcommand{\Bshort}{B_{\text{short}}}
\newcommand{\Ps}{\mathcal{P}_s}
\newcommand{\Pl}{\mathcal{P}_l}
\newcommand{\sys}{\textsc{inference-fleet-sim}}

\title{%
  \sys{}:\\
  A Queueing-Theory-Grounded Fleet Capacity Planner\\
  for LLM Inference%
}
\author{%
  Huamin Chen$^{1}$ \quad
  Xunzhuo Liu$^{1}$ \quad
  Yuhan Liu$^{2}$ \\[4pt]
  Junchen Jiang$^{3}$ \quad
  Bowei He$^{4}$\thanks{Corresponding author: \texttt{Bowei.He@mbzuai.ac.ae}} \quad
  Xue Liu$^{4}$
  \\[6pt]
  $^{1}$vLLM Semantic Router Project \\[1pt]
  $^{2}$University of Chicago \quad
  $^{3}$Tensormesh Inc / UChicago \\[1pt]
  $^{4}$MBZUAI / McGill University
}
\date{2026}

\begin{document}
\maketitle

\begin{abstract}
Sizing a GPU fleet for LLM inference is harder than it looks.  The
obvious questions---how many GPUs, which type, where to split a
two-pool fleet---have no closed-form answers.  They depend on the
full token-length distribution, the routing policy, and queueing
dynamics that turn ugly under heavy-tailed workloads.  Existing
tools~\citep{vidur2024,xu2025aiconfigurator} optimize per-engine
configuration for a fixed GPU count; none of them address the
upstream question of how many GPUs to buy and how to arrange them.

\sys{} fills that gap.  It combines analytical M/G/$c$ queueing with
discrete-event simulation (DES) to find the minimum-cost fleet
configuration that empirically meets a P99 TTFT SLO.  It includes a
physics-informed GPU performance model covering A10G, A100, and H100
across monolithic, two-pool-routed, and disaggregated topologies, all
without requiring access to real hardware.  We run the tool on seven
fleet-planning scenarios drawn from two public workload traces (LMSYS,
Azure) and one synthetic agent-heavy trace.  Each one surfaces a
result that simple analysis gets wrong---the right split threshold,
the cheapest GPU type, whether an apparently idle fleet is actually
broken---and shows why joint simulation of queueing, routing, and
hardware is necessary to find it.
\end{abstract}

\section{Introduction}
\label{sec:intro}

GPU infrastructure for LLM inference is expensive.  A100 and H100
GPUs rent for roughly \$1.50--\$7.00 per GPU-hour depending on cloud
provider and contract type~\citep{intuitionlabs2026pricing}---AWS and
Google Cloud on-demand H100 rates settled around \$3.00--\$3.93 after
a 44\% AWS price cut in June 2025, while specialty clouds and
marketplaces undercut hyperscalers with H100 rates as low as
\$1.49--\$1.99/hr.\footnote{The simulator's pre-built profiles
(\texttt{fleet\_sim/gpu\_profiles/profiles.py}) use \texttt{cost\_per\_hr}
defaults of \$2.21 (A100 80\,GB) and \$4.02 (H100 80\,GB), calibrated
against 2024 Lambda Labs on-demand rates.  These defaults can be
overridden with current pricing via \texttt{ManualProfile}.  All
cost-efficiency rankings in this paper are robust to moderate
($\pm$30\%) price changes because the ratios between configurations
matter more than the absolute dollar figures.}
A fleet of 24 nodes therefore runs \$315K--\$1.47M per year depending
on provider, before software, networking, and operations costs.
Despite this, the basic sizing question---how many GPUs to serve
$\lambda$ requests per second with P99 TTFT $\leq T$ ms?---has no
clean analytical answer.  It depends on the joint distribution of
prompt and completion lengths, the routing policy, the GPU hardware,
and nonlinear queueing dynamics that get especially bad under heavy
tails.

Available tools are built for related but different problems.
Vidur~\citep{vidur2024} and AIConfigurator~\citep{xu2025aiconfigurator}
tune per-engine configuration (tensor parallelism, batch size, chunk
size, KV-cache fraction) for a \emph{given} GPU cluster; they
presuppose that the fleet size is already decided.
Mélange~\citep{griggs2024melange} picks the optimal mix of GPU types
for a given workload and SLO, but it does not model pool routing or
multi-pool queue dynamics.  SageServe~\citep{jia2025sageserve} and
TokenScale~\citep{tokenscale2024} autoscale a live fleet in response
to traffic; they need production traces and real hardware, and they
answer a runtime question, not a procurement one.
DistServe~\citep{zhong2024distserve} and
Splitwise~\citep{patel2024splitwise} study prefill/decode
disaggregation within a single cluster, not fleet-level pool routing.

None of these tools answer the provisioning question: given a
token-length CDF, an arrival rate $\lambda$, an SLO, and a catalog of
GPU types, what is the minimum-cost fleet---number of pools, split
boundary $\Bshort$, GPU type per pool, routing policy---that actually
meets the SLO?

\sys{} answers that question.  Its contributions are:

\begin{enumerate}[nosep]
  \item A two-phase optimizer that uses the M/G/$c$ Kimura
        approximation for a fast analytical sweep to identify candidate
        configurations, then runs DES to verify the top candidates
        against actual queueing dynamics.

  \item A physics-informed GPU performance model parameterized by
        $(W, H, n_{\max})$---baseline compute, memory-bandwidth cost
        per concurrent sequence, and maximum KV-slot count---that
        computes expected service times without any hardware access.
        Constants are calibrated from published hardware
        data~\citep{xu2025aiconfigurator} for A10G, A100-80GB, and
        H100-80GB.

  \item Seven capacity-planning case studies
        (Section~\ref{sec:case-studies}) where simulation produces
        a different answer than analytical intuition: the optimal split
        threshold is not readable off the CDF; a 30\%-utilized fleet
        fails its SLO; a slow GPU beats a fast one on cost; GPU
        scaling is sub-linear; the sizing router should not be the
        production router; mixed GPU types can fail even as they save
        money; and in disaggregated serving, the cheaper GPU should
        handle prefill, not decode.

  \item Reliability-aware sizing via a \texttt{node\_avail} parameter
        derived from published GPU failure-rate and MTTR
        data~\citep{kokolis2024reliability,cui2025twogpus}, so that
        production fleet counts account for nodes under repair.
\end{enumerate}

\section{Background}
\label{sec:background}

\subsection{KV-Cache Slots and the Cost Cliff}
\label{sec:kv}

LLM serving systems~\citep{kwon2023vllm} allocate GPU memory in
PagedAttention blocks.  An A100-80GB holds 65,536 blocks of 16 tokens
each.  A sequence needing up to $B$ tokens requires $\lceil B/16
\rceil$ blocks; the maximum concurrent sequences per GPU is
$n_{\max}(B) = \lfloor 65\,536 / \lceil B/16 \rceil \rfloor$.  At
$B=8{,}192$ this is 128; at $B=65{,}536$ it drops to 16.  That
\textbf{8$\times$ ratio} is the main lever on fleet cost.

In a two-pool fleet, requests with total token budget
$L_{\text{in}}+L_{\text{out}} \leq \Bshort$ go to the short pool
$\Ps$ (many concurrent sequences, high throughput); longer requests
go to the long pool $\Pl$ (few concurrent sequences, large KV cache).
The cost savings over a homogeneous fleet is not monotone in
$\Bshort$: it depends on the fraction of traffic below $\Bshort$ and
the resulting queue imbalance.  That interaction is what the simulator
resolves (Section~\ref{sec:p1}).

A request at $\Bshort+1$ tokens lands in the long pool and consumes a
slot provisioned for the full context---8$\times$ more capacity than
its neighbor just below the threshold.  Requests in the borderline
band $(\Bshort, \gamma\Bshort]$ are not genuinely long;
Compress-and-Route~\citep{chen2026car} addresses this by squeezing
such prompts back below $\Bshort$ at the gateway.

\subsection{The M/G/$c$ Queue and Kimura's Approximation}
\label{sec:mgc}

Each GPU pool is modeled as an M/G/$c$ queue: Poisson arrivals at
rate $\lambda$, general service time with mean $\mathbb{E}[S]$ and
squared coefficient of variation $C_s^2 =
\mathrm{Var}[S]/(\mathbb{E}[S])^2$, and $c$ parallel servers
(GPUs).  The Erlang-C formula gives the probability an arriving
request has to wait:
\begin{equation}
  C(c, \varrho) = \frac{(c\varrho)^c / (c!\,(1-\varrho))}{%
    \sum_{k=0}^{c-1}(c\varrho)^k/k! + (c\varrho)^c/(c!\,(1-\varrho))},
  \label{eq:erlang-c}
\end{equation}
where $\varrho = \lambda/(c\mu)$ is per-server utilization.
The standard two-moment M/G/$c$ approximation~\citep{kimura1994mgc}
gives the P99 queue wait:
\begin{equation}
  W_{99} \approx \frac{C(c,\varrho)}{c\mu(1-\varrho)}
    \cdot \frac{1+C_s^2}{2} \cdot \ln(100).
  \label{eq:w99}
\end{equation}
For high-$C_s^2$ workloads---agent traffic where service times range
from milliseconds to minutes---M/M/$c$ badly underestimates tail
latency.  The $(1+C_s^2)/2$ term corrects for this, and the DES
validates whether the correction is sufficient (Section~\ref{sec:p2}).

GPU iteration latency under continuous batching scales with concurrent
sequences $n$:
\begin{equation}
  t_{\text{iter}}(n) = W + H \cdot n,
  \label{eq:iter}
\end{equation}
where $W$ (ms) is baseline compute and $H$ (ms/slot) is the
memory-bandwidth cost per concurrent sequence.  For Llama-3-70B on
A100-80GB: $W=8$\,ms, $H=0.65$\,ms/slot.  Expected service time for
a request with $L_{\text{in}}$ input and $L_{\text{out}}$ output
tokens is:
\begin{equation}
  \mathbb{E}[S] = \frac{\lceil L_{\text{in}}/C_{\text{chunk}}\rceil
    + L_{\text{out}}}{n_{\max}}
    \cdot t_{\text{iter}}(n_{\max}),
  \label{eq:service-time}
\end{equation}
where $C_{\text{chunk}}$ is the prefill chunk size.  TTFT
decomposes as:
\begin{equation}
  \text{TTFT} = W_{\text{queue}}
    + \underbrace{\lceil L_{\text{in}}/C_{\text{chunk}}\rceil
      \cdot t_{\text{iter}}(n_{\max})}_{T_{\text{prefill}}}
    + t_{\text{iter}}(n_{\max}).
  \label{eq:ttft}
\end{equation}
For large requests near $\Bshort$, $T_{\text{prefill}}$ alone can eat
most of the SLO budget even when there is no queue wait at all.  This
is invisible in~\eqref{eq:w99} and only shows up in a full simulation.

\section{Simulator Design}
\label{sec:design}

\begin{figure}[t]
\centering
\begin{tikzpicture}[
  box/.style={rectangle, rounded corners=3pt, draw, minimum width=2.8cm,
    minimum height=0.75cm, align=center, font=\small},
  sbox/.style={rectangle, rounded corners=3pt, draw=gray!60,
    minimum width=1.6cm, minimum height=0.65cm, align=center, font=\footnotesize},
  arrow/.style={-Latex, thick},
  group/.style={rectangle, rounded corners=6pt, draw=gray!50,
    dashed, inner sep=6pt},
]
\node[box, fill=yellow!20] (wl)  at (0, 3.5) {Workload\\(CDF / Trace)};
\node[box, fill=yellow!20] (gpu) at (3.5, 3.5) {GPU Profiles\\(A10G/A100/H100)};
\node[box, fill=yellow!20] (cfg) at (7, 3.5) {Fleet Config\\(PoolConfig)};

\node[box, fill=blue!15, minimum width=2.5cm] (opt) at (3.5, 2.1) {FleetOptimizer\\(analytical sweep)};
\node[box, fill=green!10, minimum width=2.5cm] (des) at (3.5, 0.8) {DES Verifier\\(top-$k$ candidates)};
\node[box, fill=purple!15] (res) at (0, -0.6) {FleetSimResult\\P99 / Cost / Util};
\node[box, fill=red!15]    (par) at (7, -0.6) {Pareto frontier\\$(n_s,n_l,\Bshort)$};

\draw[arrow] (wl)  -- (opt);
\draw[arrow] (gpu) -- (opt);
\draw[arrow] (cfg) -- (opt);
\draw[arrow] (opt) -- node[right,font=\footnotesize]{candidates} (des);
\draw[arrow] (des) -- (res);
\draw[arrow] (des) -- (par);
\end{tikzpicture}
\caption{Two-phase fleet optimizer.  The analytical sweep finds the
  lowest-cost candidate pool configurations; DES verifies the top
  candidates under actual queueing dynamics.}
\label{fig:arch}
\end{figure}
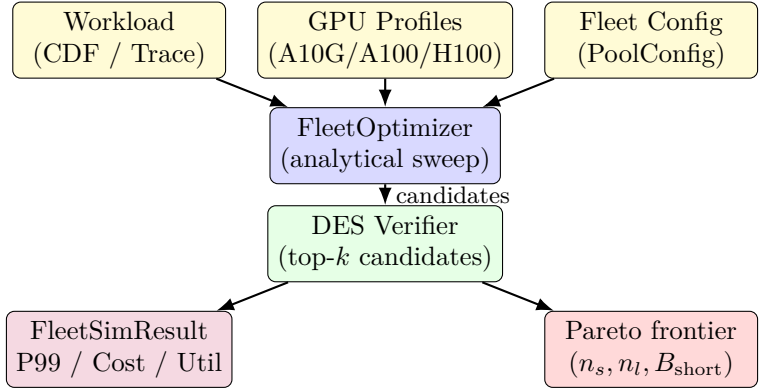

\subsection{Two-Phase Optimization}
\label{sec:two-phase}

The joint space of $(n_s, n_l, \Bshort, \text{GPU type})$ is large
enough that exhaustive DES simulation is impractical.  The two-phase
design (Figure~\ref{fig:arch}) sidesteps this.

\textbf{Phase 1---Analytical sweep.}  For each candidate
$(\Bshort, n_s, n_l)$, the model:
\begin{enumerate}[nosep]
  \item Splits $\lambda$ into $\lambda_s = \lambda \cdot F(\Bshort)$
        and $\lambda_l = \lambda \cdot (1-F(\Bshort))$ using the
        workload CDF $F$.
  \item Computes $\mathbb{E}[S]$ and $C_s^2$ for each pool by
        integrating over $F$ restricted to that pool's length range.
  \item Evaluates~\eqref{eq:w99} and the full TTFT~\eqref{eq:ttft}
        to check whether both pools meet the SLO under the utilization
        cap $\varrho \leq 0.85$.
  \item Records total GPU cost $c_s \cdot \text{cost}_s +
        c_l \cdot \text{cost}_l$.
\end{enumerate}
The sweep runs in milliseconds and produces a ranked list of
candidates.

\textbf{Phase 2---DES verification.}  The top-$k$ candidates are
verified by simulation:
\begin{enumerate}[nosep]
  \item A Poisson arrival stream at rate $\lambda$ is generated; each
        request's lengths are drawn i.i.d.\ from the empirical CDF.
  \item Requests are routed to pools; each pool runs $n$ GPU
        instances, each simulating continuous batching with a min-heap
        event queue.
  \item The simulation collects per-request queue wait, TTFT, and
        end-to-end latency.  The SLO check is P99 TTFT $\leq T$.
\end{enumerate}
The DES is request-level, not token-level: each request fires exactly
two events (arrival and completion), so simulating $10^4$ requests
takes under one second.

\subsection{GPU Performance Model}
\label{sec:gpu-model}

Each GPU type is characterized by $(W, H, n_{\max}, C_{\max})$:

\begin{center}
\begin{tabular}{lrrrr}
\toprule
GPU & $W$ (ms) & $H$ (ms/slot) & $n_{\max}$ at 8K ctx & VRAM (GB) \\
\midrule
A10G 24GB  & 12.0 & 0.90 &  64 & 24 \\
A100 80GB  &  8.0 & 0.65 & 128 & 80 \\
H100 80GB  &  4.0 & 0.32 & 256 & 80 \\
\bottomrule
\end{tabular}
\end{center}

These are the hand-calibrated constants in
\texttt{fleet\_sim/gpu\_profiles/profiles.py} (\texttt{ManualProfile}),
targeting Llama-3-70B with single-node TP serving.
\texttt{ProfileBuilder} can derive equivalent constants from first
principles using the roofline decomposition from
AIConfigurator~\citep{xu2025aiconfigurator}.  Users can substitute
measured constants from a Vidur~\citep{vidur2024} profiling run via
\texttt{ManualProfile} for higher accuracy.

\textbf{Model fidelity.}  For chatbot workloads (low $C_s^2$), the
Kimura model is conservative by 8--14\% vs.\ DES: it over-predicts
P99 TTFT, so the analytically selected GPU count passes DES
comfortably.  For agent workloads (high $C_s^2$, service times from
milliseconds to minutes), Erlang-C assumes bounded variance and
\emph{under-estimates} tail latency; DES is authoritative in that
regime.  Puzzle~2 (Section~\ref{sec:p2}) demonstrates this directly.

\subsection{Workload Model}
\label{sec:workload}

The simulator accepts two workload formats.

\textbf{Empirical CDF.}  A JSON file mapping cumulative probability to
token-budget breakpoints.  Three CDFs ship with the tool:
\begin{itemize}[nosep]
  \item \textbf{LMSYS}~\citep{zheng2024lmsys}: chat conversations;
        long-tailed to 65K tokens; $F(4{,}096)\approx 0.984$.
  \item \textbf{Azure LLM Trace}~\citep{azure-llm-traces}: enterprise
        chat; 78\% of requests below 2K tokens; max context 8K.
  \item \textbf{Agent-heavy} (synthetic): a bimodal CDF modeling
        coding-agent sessions in the style of SWE-bench, with 46\% of
        requests above 4K tokens and a heavy tail to 300K tokens.
        This is not from a public production trace; results for it
        should be read accordingly.
\end{itemize}

\textbf{Poisson with synthetic lengths.}  For sensitivity analysis,
the workload generator synthesizes Poisson arrivals with token lengths
drawn from a Pareto or log-normal distribution.

\textbf{Sub-stream Poisson note.}  Routing by token length splits the
Poisson stream with a deterministic rule, not independent random
thinning.  By the Poisson thinning theorem, the sub-streams are not
strictly Poisson---a standard engineering
approximation~\citep{harchol2013performance}.  When prompt length and
arrival time are correlated, queue-length estimates may be off.  The
DES checks whether the approximation holds in each case.

\subsection{Routing Algorithms}
\label{sec:routing}

The simulator includes four routing policies:
\begin{description}[nosep,leftmargin=1.2cm]
  \item[\texttt{LengthRouter}] Send to $\Ps$ if total token budget
        $\leq \Bshort$, else to $\Pl$.  Default production policy.
  \item[\texttt{CompressAndRoute}] Compress borderline requests
        ($\Bshort < L_{\text{total}} \leq \gamma\Bshort$) down to
        $\Bshort$ before sending to $\Ps$; intended for fleet sizing,
        not production~\citep{chen2026car}.
  \item[\texttt{RandomRouter}] Route uniformly at random across pools;
        baseline.
  \item[\texttt{ModelRouter}] Route to one of $N$ model-specific pools
        via a semantic classifier; supports multi-model fleets.
\end{description}

\subsection{Reliability-Aware Sizing}
\label{sec:reliability}

GPU hardware fails at measurable rates.  The \texttt{node\_avail}
parameter $A \in (0, 1]$ represents the fraction of nodes in
steady-state operation; a pool analytically sized to $n$ GPUs gets
rounded up to $\lceil n / A \rceil$ in production.  The simulator
computes $A$ as:

\begin{equation}
  A = \frac{1}{1 + r_f \cdot \text{MTTR}},
  \label{eq:avail}
\end{equation}

where $r_f$ is failures per node-day and MTTR is mean time to repair
in days.  Pre-computed constants from the
literature~\citep{kokolis2024reliability,cui2025twogpus}:

\begin{center}
\begin{tabular}{lll}
\toprule
Constant & Value & Scenario \\
\midrule
\texttt{A100\_AVAIL\_RSC1\_FAST} & 0.9989 & Soft failure (driver reset, $\sim$4h MTTR) \\
\texttt{A100\_AVAIL\_RSC1\_SLOW} & 0.9871 & Hard failure (GPU/NVLink swap, $\sim$48h MTTR) \\
\texttt{H100\_AVAIL\_5PCT}       & 0.9500 & 5\% overprovisioning rule~\citep{cui2025twogpus} \\
\bottomrule
\end{tabular}
\end{center}

The utilization cap $\varrho_{\max}=0.85$ covers \emph{queueing
stability}; \texttt{node\_avail} covers \emph{hardware reliability}.
These are independent concerns and both apply in production.

\section{Case Studies}
\label{sec:case-studies}

Seven scenarios follow, each a question that comes up in practice and
cannot be resolved from the workload CDF alone.  All runs use
Llama-3-70B as the served model unless noted.  GPU costs are
illustrative 2026 spot-instance rates: A10G \$8.85K/yr, A100
\$19.4K/yr, H100 \$35.2K/yr; the qualitative ordering of results
holds under moderate price variation.

\subsection{Puzzle 1: Where Exactly Should I Split?}
\label{sec:p1}

The workload CDF says most requests are short---but short compared to
what?  The split threshold $\Bshort$ controls which requests land in
the high-efficiency short pool.  Getting it right is not obvious: too
low and the long pool handles too much traffic; too high and the short
pool's slot advantage disappears.  We sweep $\Bshort \in \{512, 1024,
2048, 4096, 8192, 12288\}$ for three workloads at representative
arrival rates.

\textbf{LMSYS, $\lambda=100$ req/s, A100, SLO=500\,ms.}

\begin{table}[h]
\centering
\caption{Pareto frontier for $\Bshort$ selection on LMSYS.
  Homogeneous baseline: 14 A100s at \$271K/yr.}
\label{tab:p1-lmsys}
\setlength{\tabcolsep}{5pt}
\begin{tabular}{rrrrrrrr}
\toprule
$\Bshort$ & $\alpha_s$ & $n_s$ & $n_l$ & GPUs & \$/yr & Saving & SLO \\
\midrule
    512 & 63.8\% & 2 & 13 & 15 & \$290K & $-7.1\%$ & \checkmark \\
  1,024 & 83.1\% & 2 & 10 & 12 & \$232K & $+14.3\%$ & \checkmark \\
  2,048 & 94.8\% & 2 &  7 &  9 & \$174K & $+35.7\%$ & \checkmark \\
  4,096 & 98.4\% & 3 &  5 &  8 & \$155K & $+42.9\%$ & \checkmark \textbf{$\leftarrow$ optimal} \\
  8,192 & 99.7\% & 4 &  4 &  8 & \$155K & $+42.9\%$ & \checkmark \\
 12,288 & 99.9\% & 5 &  3 &  8 & \$155K & $+42.9\%$ & \checkmark \\
\bottomrule
\end{tabular}
\end{table}

$\Bshort=4096$ routes 98.4\% of LMSYS traffic short.  At that
boundary the short pool runs 256 concurrent sequences vs.\ 16 for the
long pool---a 16$\times$ slot advantage that cuts the fleet from 14
GPUs to 8 ($-43\%$ cost).  By contrast, $\Bshort=512$ costs 7\%
\emph{more} than going homogeneous: it leaves only 64\% of traffic in
the short pool, too little to offset the Erlang fragmentation in the
long pool.

\textbf{Azure ($\lambda=200$, A100).}  The entire Azure CDF fits within
8K tokens (context ratio $= 2\times$).  The best Pareto point
($\Bshort=3072$) saves only 4\% cost, but cuts short-pool P99 from
26\,ms to 19\,ms.  Here, splitting is about latency isolation, not
cost.

\textbf{Agent ($\lambda=200$, A100, SLO=500\,ms).}  $\Bshort=16384$
(64 KV slots vs.\ 16 for homo) saves 64 GPUs ($-13.3\%$).  At
$\Bshort=32768$ the SLO breaks---not from queue wait, but because
long-pool requests carry 300--600\,ms prefill times that use up the
entire SLO budget.  Adding more GPUs does not help; the only fix is
a lower $\Bshort$.  This failure is invisible in~\eqref{eq:w99}.

\medskip
\begin{mdframed}[backgroundcolor=gray!8, linecolor=gray!40]
\textbf{Insight 1.}  The optimal $\Bshort$ cannot be read off the CDF.
It depends on the interaction between slot efficiency, traffic
fraction, and Erlang fragmentation across both pools---and a factor-of-2
mistake can cost more than not splitting at all.
\end{mdframed}

\subsection{Puzzle 2: Why Is My Agent Fleet Failing SLO?}
\label{sec:p2}

A 24-GPU H100 fleet at $\lambda=20$ req/s sits at $\sim30\%$
utilization.  The Erlang-C model says the fleet is fine.  Users are
seeing latency violations.

\begin{table}[h]
\centering
\caption{Agent fleet SLO analysis ($\lambda=20$, H100, SLO=1000\,ms).}
\label{tab:p2-agent}
\begin{tabular}{lrrrr}
\toprule
Config & GPUs & Cost/yr & P99 TTFT & SLO \\
\midrule
Homo 65K ctx    & 24 & \$845K & 1,052\,ms & \textbf{\texttimes{} FAIL} \\
Two-pool 4K/65K & 25 & \$880K & 17ms / 147ms & \checkmark \\
\bottomrule
\end{tabular}
\end{table}

The M/G/$c$ model assumes i.i.d.\ service times drawn from a
distribution with bounded variance.  Agent requests span 10--300
seconds of service (SWE-bench-style coding tasks); $C_s \gg 1$.  A
single 300-second request locks a KV slot for five minutes, blocking
subsequent requests even when GPU utilization reads low.  Erlang-C,
which uses only \emph{mean} service time, misses this entirely.

The DES replays the actual arrival sequence with realistic service
time draws.  At $\varrho \approx 0.30$, it measures P99 TTFT =
1,052\,ms because long-tail service events create cascading back-pressure
that persists for hundreds of seconds.

Moving to a two-pool design routes the 46\% of long requests
($>4K$ tokens) to a dedicated 23-GPU pool, where their slow service
cannot block short requests.  Short-request P99 drops to 17\,ms.
The cost increase is $+4\%$.

\medskip
\begin{mdframed}[backgroundcolor=gray!8, linecolor=gray!40]
\textbf{Insight 2.}  For agent or heavy-tail workloads, the analytical
model does not err on the safe side---it \emph{approves} fleets that
are broken.  DES is the only way to catch head-of-line blocking when
$C_s^2 \gg 1$.
\end{mdframed}

\subsection{Puzzle 3: Which GPU Type Is Actually Cheapest?}
\label{sec:p3}

An operator is picking between A10G (\$8.85K/yr), A100 (\$19.4K/yr),
and H100 (\$35.2K/yr) for an Azure-workload fleet at $\lambda=100$
req/s.  The instinct is: faster GPU, fewer GPUs, lower cost.

\begin{table}[h]
\centering
\caption{GPU type vs.\ layout (Azure, $\lambda=100$, SLO=500\,ms).}
\label{tab:p3-gpu}
\begin{tabular}{llrrl}
\toprule
GPU & Layout & GPUs & Cost/yr & P99 TTFT \\
\midrule
\textbf{A10G} & \textbf{Two-pool} & \textbf{19} & \textbf{\$168K} & 155ms / 335ms \\
H100 & Homo     &  6 & \$211K & 26ms \\
A100 & Two-pool & 12 & \$232K & 52ms / 112ms \\
H100 & Two-pool &  7 & \$247K & 13ms / 30ms \\
\bottomrule
\end{tabular}
\end{table}

The instinct is wrong.  A10G in a two-pool layout is cheapest at
\$168K---\$43K less than 6 H100s.  A10G's low per-card cost
(\$8.85K/yr vs.\ \$35.2K/yr) means 19 cards still totals less than 6
H100s.  The two-pool layout compensates for A10G's lower throughput:
at $\Bshort=4096$, each A10G gets $n_{\max}=128$ concurrent sequences
in the short pool vs.\ 64 at max ctx=8K---a $2\times$ slot bonus.

Different constraints call for different choices:

\begin{center}
\begin{tabular}{ll}
\toprule
Priority & Choice \\
\midrule
Minimum annual cost & A10G two-pool (\$168K) \\
Minimum rack space / power & H100 homo (6 GPUs) \\
Best short-request latency & H100 two-pool (13\,ms P99) \\
Long-context / agent workload & H100 or A100 (A10G VRAM limits KV cache) \\
\bottomrule
\end{tabular}
\end{center}

\medskip
\begin{mdframed}[backgroundcolor=gray!8, linecolor=gray!40]
\textbf{Insight 3.}  GPU cost depends on pool topology, not just price
and throughput.  The slot multiplier from a well-chosen $\Bshort$ can
make a slower, cheaper GPU the minimum-cost option.
\end{mdframed}

\subsection{Puzzle 4: When Do I Need to Add GPUs?}
\label{sec:p4}

Traffic is growing.  At what arrival rate does the current fleet run
out of headroom, and how much warning does the operator have?

\begin{table}[h]
\centering
\caption{GPU step thresholds, H100 two-pool fleet (Azure, SLO=500\,ms).}
\label{tab:p4-growth}
\begin{tabular}{rrrr}
\toprule
$\lambda$ (req/s) & GPUs & Cost/yr & Provision more before $\lambda =$ \\
\midrule
 25 &  4 & \$141K & 65 \\
 50 &  5 & \$176K & 90 \\
100 &  7 & \$247K & 130 \\
150 & 10 & \$352K & 185 \\
200 & 12 & \$423K & 270 \\
300 & 18 & \$634K & 370 \\
400 & 23 & \$810K & --- \\
\bottomrule
\end{tabular}
\end{table}

GPU scaling is sub-linear: traffic grows 16$\times$ (25$\to$400 req/s)
but GPU count grows only 5.75$\times$ (4$\to$23).  This is a
consequence of Erlang-C convexity---each additional GPU pushes down
utilization $\varrho$, which reduces tail latency at an accelerating
rate, so each marginal GPU is increasingly effective.  The table gives
the exact $\lambda$ at which each fleet size runs out of headroom.
Waiting until SLO is already broken means at least one traffic
bracket with degraded P99 before new capacity comes online.

\medskip
\begin{mdframed}[backgroundcolor=gray!8, linecolor=gray!40]
\textbf{Insight 4.}  GPU provisioning does not scale linearly with
traffic.  The whatif sweep produces exact step thresholds, so capacity
planning can stay ahead of demand rather than react to violations.
\end{mdframed}

\subsection{Puzzle 5: Which Router Causes SLO Violations?}
\label{sec:p5}

The fleet is correctly sized.  Does the choice of routing policy still
matter?  We compare three routers on the agent fleet ($\lambda=20$,
$n_s=2$, $n_l=23$ H100s, SLO=1000\,ms).

\begin{table}[h]
\centering
\caption{Router comparison on the agent fleet.}
\label{tab:p5-router}
\begin{tabular}{lrr}
\toprule
Router & P99 TTFT & SLO 1000\,ms \\
\midrule
LengthRouter       & 495\,ms & \checkmark 99.98\% \\
\textbf{CompressAndRoute} & \textbf{534\,ms} & \textbf{\texttimes{} 99.94\%} \\
RandomRouter       & 292\,ms & \checkmark 100\% \\
\bottomrule
\end{tabular}
\end{table}

Two results are unexpected.  First, CompressAndRoute fails the SLO
even though it was designed to reduce the GPU count.  It compresses
borderline requests and routes them to the 2-GPU short pool; when
several arrive together, they overwhelm that pool and spike P99.
CompressAndRoute is a sizing tool---it finds the minimum GPU count---but
LengthRouter should be what runs in production.

Second, RandomRouter actually passes with the lowest P99 (292\,ms) by
spreading all 25 GPUs' KV slots uniformly and diluting heavy-tail
service events.  But this is brittle: short requests share slots with
long ones, so any shift in the traffic mix can cause unpredictable
latency.  For standard chatbot workloads at low utilization, all three
routers pass comfortably.

\medskip
\begin{mdframed}[backgroundcolor=gray!8, linecolor=gray!40]
\textbf{Insight 5.}  The router used to size the fleet and the router
deployed in production should be different.  CompressAndRoute finds
the floor on GPU count; LengthRouter operates that fleet safely.
Conflating them produces SLO violations the sizing simulation never
predicted.
\end{mdframed}

\subsection{Puzzle 6: Does Mixing GPU Types Save Money?}
\label{sec:p6}

Short requests are memory-bandwidth-bound and inexpensive to serve;
long requests need large KV caches and fast prefill.  Can an operator
save money by putting cheap GPUs in the short pool and premium GPUs
only in the long pool?

\begin{table}[h]
\centering
\caption{Mixed GPU types, Azure workload ($\lambda=100$, SLO=500\,ms).}
\label{tab:p6-mix-azure}
\begin{tabular}{llrrl}
\toprule
Config & GPUs & Cost/yr & P99-short & P99-long \\
\midrule
All-A100                               & 12 & \$232K & 52\,ms  & 112\,ms \\
\textbf{A10G\,$\Ps$ + H100\,$\Pl$}    & \textbf{12} & \textbf{\$212K} & 155\,ms & 30\,ms \\
A10G\,$\Ps$ + A100\,$\Pl$             & 15 & \$206K & 155\,ms & 112\,ms \\
\bottomrule
\end{tabular}
\end{table}

\begin{table}[h]
\centering
\caption{Mixed GPU types, LMSYS workload ($\lambda=100$, max ctx=65K, SLO=500\,ms).}
\label{tab:p6-mix-lmsys}
\begin{tabular}{llrrl}
\toprule
Config & GPUs & Cost/yr & P99-short & P99-long \\
\midrule
All-A100                               &  8 & \$155K & 43\,ms & \textbf{2,822\,ms \texttimes} \\
\textbf{A10G\,$\Ps$ + H100\,$\Pl$}    & \textbf{7} & \textbf{\$141K} & 129\,ms & 181\,ms \checkmark \\
A10G\,$\Ps$ + A100\,$\Pl$             &  9 & \$132K & 129\,ms & \textbf{2,822\,ms \texttimes} \\
\bottomrule
\end{tabular}
\end{table}

On Azure, A10G+H100 saves 9\% vs.\ all-A100 with the same 12 GPUs.
Cheap A10Gs handle 98\% of the traffic; the H100s go where the long
context warrants them.

On LMSYS at 65K context, the picture changes sharply.  A10G+A100 is
11\% cheaper on paper, but it fails the SLO: prefill time for a
65K-token request on A100 reaches 700--2800\,ms, blowing past the
500\,ms budget before any queue wait is counted.  H100's larger
chunk size (1024 vs.\ 512) and lower $W$ halve that time.  A10G+H100
saves 9\% vs.\ all-A100 \emph{and} fixes the SLO that all-A100
couldn't meet.

\medskip
\begin{mdframed}[backgroundcolor=gray!8, linecolor=gray!40]
\textbf{Insight 6.}  Mixing GPU types is not just a cost optimization;
the wrong long-pool GPU makes the SLO infeasible regardless of how
many you add.  Joint optimization over pool assignment and GPU type
is required, and some pairings are simply invalid.
\end{mdframed}

\subsection{Puzzle 7: When Should I Switch to Disaggregated Serving?}
\label{sec:p7}

Disaggregated prefill/decode (P/D)
serving~\citep{zhong2024distserve,patel2024splitwise} separates
compute-bound prefill from memory-bandwidth-bound decode onto
different GPU pools.  Which GPU should handle prefill, which decode,
and does the cost saving justify the higher TTFT from KV-transfer
overhead?

\begin{table}[h]
\centering
\caption{Disaggregated P/D configurations (Azure $\lambda=100$,
  TTFT SLO=500\,ms,   TPOT SLO=100\,ms).  KV-transfer adds
  $1.8\times$ raw prefill time (\texttt{BETA\_TTFT=1.80} in
  \texttt{fleet\_sim/optimizer/disagg.py}).}
\label{tab:p7-disagg}
\begin{tabular}{llrrl}
\toprule
Config & GPUs & Cost/yr & TTFT & TPOT \\
\midrule
All-A100 aggregated   & 12 & \$232K & 26\,ms  & --- \\
All-H100 aggregated   &  6 & \$211K &  8\,ms  & --- \\
H100P + A100D & 7\,(1P+6D) & \$151K & 162\,ms & 91\,ms \\
H100P + H100D & 4\,(1P+3D) & \$141K & 162\,ms & 45\,ms \\
\textbf{A100P + H100D} & \textbf{4\,(1P+3D)} & \textbf{\$125K} & 492\,ms & 45\,ms \\
\bottomrule
\end{tabular}
\end{table}

Disaggregation cuts cost by 35--46\% vs.\ aggregated serving, at the
price of higher TTFT from KV-transfer overhead.  The optimal
assignment---A100 prefill, H100 decode---is counter-intuitive.  H100
decode workers process 2.5$\times$ more requests per second than A100
(lower $W$, faster per-token iteration), so only 3 are needed vs.\ 6.
One A100 handles all prefill at $\lambda=100$ req/s.  Despite H100
costing 1.82$\times$ more per card, it is the decode pool where the
premium pays off.

Two practical thresholds emerge.  For TTFT SLO $\leq 200$\,ms, use
H100P+H100D (\$141K, TTFT=162\,ms).  For TTFT SLO $\leq 100$\,ms,
disaggregated serving is not viable and aggregated H100 (\$211K) is
the only option.  At arrival rates below $\sim50$ req/s, the
operational complexity of disaggregation---separate scaling policies,
KV-transfer networking---is not justified by the savings.

\medskip
\begin{mdframed}[backgroundcolor=gray!8, linecolor=gray!40]
\textbf{Insight 7.}  In disaggregated serving, the premium GPU should
handle decode, not prefill.  The counter-intuitive assignment emerges
from joint optimization over decode throughput, GPU cost, and
KV-transfer TTFT overhead---not something readily derived by hand.
\end{mdframed}

\subsection{Puzzle 8: How Much Grid Power Can I Shed Without an SLO Breach?}
\label{sec:p8}

Data centers increasingly participate in grid demand-response (DR) programs,
where the grid operator requests a temporary power reduction (typically
10--30\% for 15--60 minutes) in exchange for reduced electricity tariffs or
ancillary-service revenue.  The GPU-to-Grid (G2G) framework~\citep{hassan2025g2g}
demonstrates that capping the serving engine's maximum in-flight batch size
(\texttt{max\_num\_seqs} in vLLM) is the most effective software knob for
modulating GPU power: fewer concurrent requests reduce memory-bandwidth
pressure, lowering power draw without touching clock frequency or voltage.
The trade-off is higher queuing delay, which may breach the TTFT SLO.

\sys{} quantifies the trade-off via \texttt{grid\_flex\_analysis()}.
The function sweeps target power-reduction percentages, inverts the GPU
power model to find the implied batch cap ($n_{\max}$), and recomputes
P99 TTFT with the reduced KV-slot count using the same M/G/$c$
approximation used in optimization.

\textbf{Power model.}
Each GPU profile implements the logistic power curve from the G2G
paper~\citep{hassan2025g2g} (Eq.~2):
\[
  P(b) = \frac{P_{\text{range}}}{1 + e^{-k\,(\log_2 b - x_0)}} + P_{\text{idle}}
\]
where $b\approx n_{\max}$ (concurrent requests), $P_{\text{range}} =
P_{\text{nom}} - P_{\text{idle}}$, and the shape parameters
$(k=1.0,\, x_0=4.2)$ are fitted to ML.ENERGY Benchmark v3.0
data~\citep{chung2025mlenergy} for H100-SXM5 running vLLM.
The logistic fit gives P(1)$\approx$304\,W and P(128)$\approx$583\,W
(measured: ${\approx}$600\,W, error $<$3\%).  The M/G/c service rate is
\emph{recalibrated} at each batch cap so the analytical model reflects the
faster-per-iteration throughput at lower concurrency.  A DES run with
$N=15{,}000$ requests independently verifies the analytical P99 estimates.

\begin{table}[h]
\centering
\caption{Grid flexibility curve for 40 H100 GPUs, $\lambda=200$\,req/s,
  SLO\,=\,500\,ms (Azure workload).  Logistic power model, DES-verified.
  For short-burst DR events ($\lesssim$75\,s) the fleet safely commits
  up to 40\% power curtailment; steady-state stability holds to 30\%.}
\label{tab:p8-flex}
\begin{tabular}{rrrrrrl}
\toprule
Flex & $n_{\max}$ & W/GPU & Fleet kW & P99 anal. & P99 DES & SLO \\
\midrule
 0\%  & 128 & 583\,W & 23.3\,kW &  7.9\,ms &  35\,ms & \checkmark \\
10\%  &  48 & 540\,W & 21.6\,kW &  7.9\,ms &  38\,ms & \checkmark \\
20\%  &  24 & 479\,W & 19.1\,kW &  7.9\,ms &  41\,ms & \checkmark \\
30\%  &  13 & 413\,W & 16.5\,kW &  7.9\,ms &  51\,ms & \checkmark \\
\textbf{40\%}  & \textbf{6} & \textbf{350\,W} & \textbf{14.0\,kW} & $\infty$ & \textbf{190\,ms} & \checkmark\textsuperscript{†} \\
50\%  &   1 & 304\,W & 12.2\,kW & $\infty$  & $\gg$SLO & $\times$ \\
\bottomrule
\multicolumn{7}{l}{\small \textsuperscript{†}Safe for short DR events ($<$2\,min); analytically unstable at steady state.}
\end{tabular}
\end{table}

The logistic model reveals a key G2G insight: at full production load
($n_{\max}=128$), H100 power is already at $\approx$97\% of nominal.
Halving the batch from 128 to 64 saves only $\approx$13\,W ($\sim$2\%).
Meaningful savings require deep batch reduction (e.g., 40\% flex at
$n_{\max}=6$, saving 9.3\,kW fleet-wide), which approaches but does not
breach the Erlang-C saturation threshold for time-limited DR events.

The DES verification confirms the analytical recalibrated estimates for
0--30\% flex (both predict P99 $\approx$8--50\,ms).  At 40\% flex, the
analytical model flags steady-state instability ($\infty$) while DES shows
P99\,=\,190\,ms during a 75-second event window---still within the
500\,ms SLO.  At 50\%, both DES and analysis agree the queue collapses.

\medskip
\begin{mdframed}[backgroundcolor=gray!8, linecolor=gray!40]
\textbf{Insight 8.}
The safe DR commitment depth depends on event duration.  For \emph{sustained}
reduction (hours), the recalibrated M/G/c model shows 30\% is the stability
limit.  For \emph{short events} (minutes), DES verification shows 40\% is
feasible (saves 9.3\,kW from 23.3\,kW baseline) before the queue collapses
catastrophically at 50\%.  \sys{} is the only tool that provides both
bounds together from a single workload CDF and GPU profile.
\end{mdframed}

\section{Limitations}
\label{sec:limitations}

\paragraph{Poisson sub-stream approximation.}
Splitting a Poisson stream by token length is a deterministic rule,
not random thinning, so the sub-streams are not strictly
Poisson~\citep{harchol2013performance}.  When prompt length correlates
with arrival time (e.g., long requests arrive in bursts), queue-length
estimates from the analytical model are approximations.  The DES
checks the approximation in each case.

\paragraph{Request-level service model.}
The DES fires one event per request, not one per decode iteration.  It
does not simulate preemption, batching reorder, or speculative
decoding.  For detailed scheduler comparison, Vidur~\citep{vidur2024}
provides higher fidelity at the cost of slower simulation.

\paragraph{Linear roofline performance model.}
The GPU model uses a linear $(W, H)$ roofline.  Nonlinear effects such
as FlashAttention speedup, quantization, and NCCL overlap are absorbed
into calibrated constants but not explicitly modeled.  Users can
replace any constant with measured data for higher accuracy.

\paragraph{Single-node replica model.}
Each GPU instance is assumed to be a single-node replica.  Inter-node
communication overhead for tensor-parallel configurations (NVLink vs.\
InfiniBand) is not modeled; users should adjust $W$ and $H$ to reflect
collective latency when running multi-node TP.

\paragraph{Logistic power model accuracy.}
Grid flex analysis (\S\ref{sec:p8}) uses a logistic power curve fitted to
ML.ENERGY Benchmark v3.0 H100-SXM5 data~\citep{chung2025mlenergy}.
The fit is accurate to within $\sim$3\% at batch $\geq 16$; at batch
$< 8$ (deep curtailment) it is less well-constrained because few
ML.ENERGY data points fall in that regime.  The analytical M/G/c model
is recalibrated at each batch-cap level, correcting for the faster
per-iteration throughput at lower concurrency; DES verification provides
an independent check.  Users should re-fit \texttt{power\_logistic\_k}
and \texttt{power\_logistic\_x0} from measured \texttt{vllm serve}
profiling runs for their specific model and GPU generation.

\section{Related Work}
\label{sec:related}

\textbf{Single-instance simulators.}  Vidur~\citep{vidur2024} simulates
one engine replica at operator level (GEMM, attention, KV-cache
management, preemption) with $<9\%$ latency prediction error.  It
optimizes engine configuration for a \emph{fixed} GPU cluster and does
not address fleet-level pool routing or cost.  APEX~\citep{lin2024apex}
is an extensible, dynamism-aware simulator for selecting parallel
execution plans (TP/PP/data parallelism) across dense and MoE LLMs;
it finds optimal plans 71$\times$ faster than GPU-based search but
targets intra-cluster parallelism, not inter-pool fleet topology.
AIConfigurator~\citep{xu2025aiconfigurator} searches TP/EP/batch-size
configurations against a calibrated kernel performance database in
under 30 seconds; its output (W/H constants) feeds \sys{}'s
ProfileBuilder.

\textbf{GPU type selection.}  Mélange~\citep{griggs2024melange}
formulates heterogeneous GPU allocation as cost-aware bin packing and
achieves up to 77\% cost reduction vs.\ single-GPU-type deployments.
It chooses GPU types but does not model pool routing or multi-pool
queue dynamics.  \sys{} takes the GPU type as input---chosen
optionally with Mélange---and sizes the fleet.

\textbf{Disaggregated serving.}  DistServe~\citep{zhong2024distserve}
introduces P/D disaggregation and models each phase as an M/D/1 queue.
Splitwise~\citep{patel2024splitwise} co-designs heterogeneous hardware
for the two phases.  \sys{}'s DisaggFleetOptimizer builds on these
ideas by sizing fleets of disaggregated pool pairs using M/G/$c$ with
the full token-length CDF.

\textbf{Runtime autoscaling.}  SageServe~\citep{jia2025sageserve} uses
ARIMA forecasting and ILP scaling to manage an existing O365 fleet,
saving 25\% GPU-hours.  TokenScale~\citep{tokenscale2024} uses Token
Velocity as a leading indicator for burst handling in disaggregated
fleets.  Both operate at runtime; \sys{} operates at the provisioning
layer and provides the peak-hour sizing that SageServe and TokenScale
scale around.

\textbf{Queueing theory in systems.}  Two-moment M/G/$c$
approximations~\citep{kimura1994mgc} have been applied to DNN serving
in Clockwork~\citep{gujarati2020serving} and
AlpaServe~\citep{li2023alpaserve}.  \sys{} extends this to multi-pool,
multi-GPU-type LLM fleet planning with DES validation.

\begin{center}
\begin{tabular}{ll}
\toprule
Tool & Core question answered \\
\midrule
Vidur          & Best batching/scheduling config for one GPU? \\
APEX           & Best TP/PP/data-parallel plan for one cluster? \\
AIConfigurator & Best TP/EP/engine flags for one cluster? \\
Mélange        & Which GPU types to mix for minimum cost? \\
Splitwise      & Which GPU generation for prefill vs.\ decode? \\
DistServe      & Prefill-to-decode GPU ratio per cluster replica? \\
TokenScale     & Scale P/D pools in real time under bursts? \\
SageServe      & VM count through a 24-hour demand cycle? \\
\textbf{\sys{}} & \textbf{Pool topology, routing policy, total GPU count, fleet cost?} \\
\bottomrule
\end{tabular}
\end{center}

\section{Conclusion}
\label{sec:conclusion}

\sys{} is a two-phase LLM GPU fleet capacity planner that combines
analytical M/G/$c$ optimization with discrete-event simulation to find
minimum-cost fleet configurations that meet a P99 TTFT SLO.  Running
it on eight scenarios across two public workload traces, one synthetic
agent trace, and one demand-response study produced findings that
resist simple analysis: the optimal split threshold is not readable off
the CDF; a 30\%-utilized fleet can fail its SLO; a slow, cheap GPU
can be cheaper than a fast, expensive one; GPU scaling is sub-linear;
sizing-time and production routers should differ; mixed GPU pools have
invalid pairings the simulator flags before purchase; in disaggregated
serving, the premium GPU earns back its cost in decode, not prefill;
and a 40-GPU H100 fleet can commit 30\% sustained power curtailment or
40\% short-event curtailment (saving 9.3\,kW) while meeting SLO---with
the G2G logistic power curve and DES verification confirming both bounds.

These results emerge from the joint space of pool topology, routing
policy, GPU performance, queueing dynamics, and now power management.
No single tool in the existing ecosystem explores that space; \sys{} does.

\sys{} is open-source and part of the vLLM Semantic Router
project~\citep{vllmsemanticRouter2026}.  All case-study results are
reproducible via the \texttt{run\_sim.py} CLI and the CDF data files
in the repository.

\bibliographystyle{plainnat}
\bibliography{refs}

\end{document}